\documentclass[12pt]{article}
\usepackage{graphics}
\usepackage{epsfig}
\usepackage{color}
\usepackage{amsmath}
\usepackage{amssymb}
\def\da{\dot{\alpha}}
\def\db{\dot{\beta}}
\def\dg{\dot{\gamma}}
\def\dd{\dot{\delta}}

\def\D{\Delta}
\def\L{\Lambda}

\def\a{\alpha}
\def\b{\beta}
\def\g{\gamma}
\def\d{\delta}
\def\e{\varepsilon}
\def\m{\mu}

\def\s{\sigma}
\def\r{\rho}
\def\l{\lambda}
\def\t{\tau}
\def\o{\omega}

\def\vt{\vartheta}
\def\mc{\mathcal}

\def\p{\partial}
\def\la{\langle}
\def\ra{\rangle}
\def\dag{\dagger}
\def\wt{\widetilde}
\def\K{\widetilde{K}}

\def\lr{\Leftrightarrow}
\numberwithin{equation}{section} 
\parskip=\medskipamount

\def\Zop{\bbbz}
\def\Rop{\bbbr}
\def\Nop{\bbbn}
\def\bbbz {{\sf Z\!\!Z}}

\def\bbbr {{\rm I\!R}}
\def\bbbn {{\rm I\!N}}
\def\RR{R-R }

\begin{document}
\thispagestyle{empty}
\addtocounter{page}{-1}
\def\thefootnote{\fnsymbol{footnote}}
\begin{flushright}
  hep-th/0304232 \\
  AEI-2003-040
\end{flushright}

\vskip 0.5cm

\begin{center}\LARGE
{\bf An alternative formulation of light-cone \\ string field theory on the plane wave}
\end{center}

\vskip 1.0cm

\begin{center}
{\large Ari Pankiewicz\footnote{E-mail address: {\tt apankie@aei.mpg.de}}}

\vskip 0.5cm

{\it $^*$ Max-Planck-Institut f\"ur Gravitationsphysik, Albert-Einstein-Institut \\
Am M\"uhlenberg 1, D-14476 Golm, Germany}
\end{center}

\vskip 1.0cm

\begin{center}
April 2003
\end{center}

\vskip 1.0cm

\begin{abstract}
\noindent
We construct a manifestly $SO(4)\times SO(4)$ invariant,
supersymmetric extension of the closed string cubic interaction vertex and dynamical
supercharges in light-cone string field theory on the plane wave space-time.
We find that the effective vertex for states built out of bosonic creation
oscillators coincides with the one previously constructed in the
$SO(8)$ formalism and conjecture that in general the two formulations
are physically equivalent.
Further evidence for this claim is obtained from the discrete
$\Zop_2$-symmetry of the plane wave and by
computing the mass-shift of the simplest
stringy state using perturbation theory.
We verify that the leading non-planar correction to
the anomalous dimension of the dual gauge theory operators is
correctly recovered.
\end{abstract}

\vfill

\setcounter{footnote}{0}
\def\thefootnote{\arabic{footnote}}
\newpage

\renewcommand{\theequation}{\thesection.\arabic{equation}}
\section{Introduction}

The maximally supersymmetric pp-wave space-time~\cite{Blau:2001ne} (henceforth referred to as the plane wave)
\begin{equation}\label{planewave}
\begin{split}
ds^2 & = 2dx^+dx^--\m^2\vec{x}^2\bigl(dx^+\bigr)^2+d\vec{x}^2\,,\\
F_5 & = 4\m dx^+\wedge\bigl(dx^1\wedge dx^2\wedge dx^3\wedge dx^4+dx^5\wedge dx^6\wedge dx^7\wedge dx^8\bigr)\,.
\end{split}
\end{equation}
provides an interesting and tractable example of a curved background with \RR flux on which many aspects of (type IIB)
string theory can be studied.
In particular, in the Green-Schwarz formalism, the world-sheet action is quadratic in light-cone gauge and hence can be
quantized exactly~\cite{Metsaev:2001bj}. The spectrum of the closed string consists of a unique massless ground-state $|v\ra$ and an
infinite tower of excited states whose masses are of order $\o_n/{\a'p^+}$~\cite{Metsaev:2002re}, where
\begin{equation}
\o_n=\sqrt{n^2+\bigl(\m\a'p^+\bigr)^2}\,,\qquad n\in\Nop\,,
\end{equation}
$p^+$ is the light-cone momentum.
An important aspect of the plane wave is that it can be obtained as a
Penrose-G\"{u}ven limit~\cite{Penrose1976,Gueven:2000ru} of $AdS_5\times S^5$~\cite{Blau:2002dy,Berenstein:2002jq}.
This allowed~\cite{Berenstein:2002jq} to derive a duality between string theory on the plane wave and a double-scaling limit of
${\mc N}=4$ $SU(N)$ super Yang-Mills (SYM) from the AdS/CFT correspondence~\cite{Maldacena:1998re,Gubser:1998bc,Witten:1998qj}.
The latter can then be studied, albeit in a special limit, beyond the supergravity approximation due to the solvability of
string theory in this background.

The Penrose limit induces the following relations between string and gauge theory parameters~\cite{Berenstein:2002jq}
 \begin{equation}\label{dict}
\frac{1}{\m}H = \D-J\,,\qquad \frac{1}{\bigl(\m\a'p^+\bigr)^2} = \frac{g^2_{\text{YM}}N}{J^2}\equiv \l'\,,\qquad
4\pi g_{\text{s}}\bigl(\m\a'p^+\bigr)^2 = \frac{J^2}{N}\equiv g_2
\end{equation}
where $\D$ is the conformal dimension and $J$ the charge of a $U(1)_R$ subgroup (corresponding to a $S^1$ inside the $S^5$ that is
singled out by the Penrose limit) of the $SU(4)_R$ $R$-symmetry of SYM. The composite (BMN) operators with large charge $J$ and conformal
dimension $\D$ whose difference $\D-J$ remains fixed in the double scaling limit
\begin{equation}
N\to\infty\quad\text{and}\quad J\to\infty\quad\text{with}\quad\frac{J^2}{N}\quad
\text{fixed}\,,\quad g_{\text{YM}}\quad\text{fixed}\,,
\end{equation}
are the field theory duals of perturbative string states~\cite{Berenstein:2002jq}. For these operators the quantity $\D-J$ is a function
of the effective coupling $\l'$~\cite{Berenstein:2002jq,Gross:2002su,Santambrogio:2002sb} and the effective genus counting parameter
$g_2^2$~\cite{Kristjansen:2002bb,Constable:2002hw} (see also~\cite{Berenstein:2002sa,Gross:2002mh}). Subsequently the gauge theory side
of this novel duality has been extensively studied, see
e.g.~\cite{Beisert:2002bb,Constable:2002vq,Gursoy:2002yy,Eynard:2002df,Janik:2002bd,Beisert:2002tn,Minahan:2002ve,
Beisert:2002ff,Klose:2003tw,Beisert:2003tq}.

String interactions in the plane wave space-time correspond to the non-planar sector (i.e. finite $g_2$) of interacting gauge theory,
cf.~\eqref{dict}. In light-cone gauge in flat space-time string interactions can be studied using the vertex operator approach. The
difficulties associated with the fact that $x^-$ is quadratic in transverse coordinates can be circumvented exploiting the
ten-dimensional Lorentz-symmetry. In the plane wave this is broken by the \RR flux, in particular there is no $J^{+-}$ generator, and the
study of interactions is more involved. On the other hand, the light-cone string field theory formalism
developed in flat space-time for the bosonic string (see e.g.~\cite{Mandelstam:1973jk,Mandelstam:1974fq,Cremmer:1974jq,Cremmer:1975ej})
and extended to the superstring in~\cite{Mandelstam:1974hk,Green:1983tc,Green:1983hw,Green:1984fu} was successfully generalized to the
plane wave for closed
strings~\cite{Spradlin:2002ar,Spradlin:2002rv,Pankiewicz:2002gs,Pankiewicz:2002tg,Klebanov:2002mp,Schwarz:2002bc,He:2002zu} and
for open strings~\cite{Chandrasekhar:2003fq,Stefanski:2003zc}.
In light-cone string field theory the cubic interaction vertex is a first order in $g_{\text{s}}$ correction to the free Hamiltonian.
In the bosonic string it is constructed by requiring continuity of string fields on the world-sheet;
in the number basis continuity is enforced mode by mode by an exponential of creation oscillators.
In the superstring additional `dynamical' constraints arise from the requirement that the superalgebra is realized in the
interacting theory; hence the dynamical supercharges that anticommute to the Hamiltonian have to acquire corrections as well.
These additional constraints on the interaction vertex (and dynamical supercharges)
are solved by introducing prefactors, polynomial in creation oscillators, acting on the exponential part of the vertex.
In the functional formalism these prefactors have the interpretation of interaction point operators~\cite{Mandelstam:1974hk,Green:1983tc}.

Light-cone string field theory has been used to perform a successful test of the duality
beyond the free string/planar gauge theory level in~\cite{Roiban:2002xr}, where
the leading non-planar correction of certain BMN
operators~\cite{Beisert:2002bb,Constable:2002vq} was reproduced from a string field
theory computation (see also~\cite{Gomis:2003kj}). For further aspects of string interactions and
the duality to SYM see
e.g.~\cite{Huang:2002wf,Chu:2002pd,Lee:2002rm,Lee:2002vz,Gomis:2002wi,Kiem:2002pb,Verlinde:2002ig,Zhou:2002mi,Vaman:2002ka,Pearson:2002zs,Bak:2002ku,Spradlin:2003bw}.
In spite of this success there remains a puzzle concerning the correct approach to light-cone string field
theory on the plane wave background. Whereas~\cite{Spradlin:2002ar,Spradlin:2002rv,Pankiewicz:2002gs,Pankiewicz:2002tg}
used the $SO(8)$ formalism to construct the cubic interaction vertex and thereby closely followed the flat space-time
approach~\cite{Green:1983tc,Green:1983hw}, an alternative and apparently different construction
(called the $SO(4)\times SO(4)$ formalism in what follows) was initiated
in~\cite{Chu:2002eu,Chu:2002wj}, where the continuity conditions were solved.
The difference in the two formulations lies in their starting point: in the $SO(8)$ formalism the vertex is built on the
state $|0\ra$ (with energy $\sim\m$) that is not the ground state of string theory on the plane wave, rather it corresponds to the
dilaton/axion fluctuation mode of plane wave supergravity~\cite{Metsaev:2002re}. Then, by construction, one has a smooth connection to
flat space-time string field theory as $\m\to 0$. On the other hand,
in the $SO(4)\times SO(4)$ setup, the vertex is built on the vacuum $|v\ra$
of plane wave string theory, which corresponds to the fluctuation mode of a mixture of the trace of the graviton and the \RR potential
on one of the two transverse $\Rop^4$'s~\cite{Metsaev:2002re}.
The presence of the \RR flux breaks the transverse $SO(8)$ symmetry of the metric to $SO(4)\times SO(4)\times\Zop_2$, where
the discrete $\Zop_2$ is a particular $SO(8)$ transformation that exchanges the two $\Rop^4$'s (cf.~\eqref{planewave}).
Based on this $\Zop_2$ symmetry it was argued in~\cite{Chu:2002eu} that only the $SO(4)\times SO(4)$ formalism
yields a vertex which respects the complete transverse symmetry. In this paper we attempt to resolve this puzzle by solving
the dynamical constraints in the $SO(4)\times SO(4)$ formalism. With the complete solution for the cubic interaction vertex and dynamical
supercharges at hand, we find strong indications that {\em both} formulations preserve
all the transverse symmetries and propose that they are physically equivalent.

This paper is organized as follows. In section~\ref{section2} we briefly review the free string
and recall the solution to the continuity conditions in the $SO(4)\times SO(4)$ formalism. In section~\ref{section3} we
present the solution of the dynamical constraints, i.e. the prefactors both of the cubic interaction vertex and dynamical
supercharges. In section~\ref{section4} we use these results to compute the leading order (for large $\m$) mass-shift of the simplest
stringy state using perturbation theory and truncation to the so called impurity-conserving channel. We conclude in section~\ref{section5}.
Several appendices are also included: in appendix~\ref{appA} we summarize our notations and conventions, appendices~\ref{appB}
and~\ref{appC} contain details on the derivation of the prefactors and appendix~\ref{appD} provides the functional expressions for
the fermionic constituents of the prefactors.

\setcounter{equation}{0}

\section{Preliminaries}\label{section2}

\subsection{The free string}

The free string in the plane wave background in
light-cone gauge is described by $x^I_r(\s_r)$ and $\vt^a_r(\s_r)$ in position space or
by $p^I_r(\s_r)$ and $\l^a_r(\s_r)$ in momentum space, where $I$ is a transverse
SO(8) vector index, $a$ is a SO(8) spinor index. The index $r=1,2,3$ denotes the $r$th
string. The bosonic part of the light-cone action in the plane wave background is~\cite{Metsaev:2001bj}
\begin{equation}
S_{\mbox{\scriptsize bos.}(r)}=\frac{e(\a_r)}{4\pi\a'}\int\,d\t\int_0^{2\pi|\a_r|}
\,d\s_r\bigl[\dot{x}_r^2-x^{\prime\,2}_r-\m^2x_r^2\bigr]\,,
\end{equation}
where $\dot{x}_r\equiv\p_{\t}x_r$, $x'_r\equiv\p_{\s_r}x_r$, $\a_r\equiv\a'p^+_r$ and
$e(\a_r)\equiv\a_r/|\alpha_r|$.
In a collision process $p^+_r$ will be positive for an incoming string and negative for an outgoing one.
The mode expansions of the fields $x_r^I(\s_r,\t)$ and $p_r^I(\s_r,\t)$ at $\t=0$ are
\begin{equation}
\begin{split}
x_r^I(\s_r)& = x_{0(r)}^I+\sqrt{2}\sum_{n=1}^{\infty}
\bigl(x_{n(r)}^I\cos\frac{n\s_r}{|\a_r|}+x_{-n(r)}^I\sin\frac{n\s_r}{|\a_r|}\bigr)\,,\\
p_r^I(\s_r) & =\frac{1}{2\pi|\a_r|}\bigl[p_{0(r)}^I+\sqrt{2}\sum_{n=1}^{\infty}
\bigl(p_{n(r)}^I\cos\frac{n\s_r}{|\a_r|}+p_{-n(r)}^I\sin\frac{n\s_r}{|\a_r|}\bigr)\bigr]\,.
\end{split}
\end{equation}
The Fourier modes can be reexpressed in terms of creation and annihilation operators as
\begin{equation}\label{xp}
x_{n(r)}^I=i\sqrt{\frac{\a'}{2\o_{n(r)}}}\bigl(a_{n(r)}^I-a_{n(r)}^{I\,\dag}\bigr)\,,\qquad
p_{n(r)}^I=\sqrt{\frac{\o_{n(r)}}{2\a'}}\bigl(a_{n(r)}^I+a_{n(r)}^{I\,\dag}\bigr)\,.
\end{equation}
Canonical quantization of the bosonic coordinates yields the usual commutation relations
\begin{equation}
[x_r^I(\s_r),p_s^J(\s_s)]=i\d^{IJ}\d_{rs}\d(\s_r-\s_s)\qquad\lr\qquad [a_{n(r)}^I,a_{m(s)}^{J\,\dag}]=\d^{IJ}\d_{nm}\d_{rs}\,.
\end{equation}
The fermionic part of the light-cone action in the plane wave is~\cite{Metsaev:2001bj}
\begin{equation}
S_{\mbox{\scriptsize ferm.}(r)}=\frac{1}{8\pi}\int\,d\t\int_0^{2\pi|\a_r|}\,d\s_r
[i(\bar{\vt}_r\dot{\vt}_r+\vt_r\dot{\bar{\vt}}_r)
-\vt_r\vt'_r+\bar{\vt}_r\bar{\vt}'_r-2\m\bar{\vt}_r\Pi\vt_r]\,,
\end{equation}
where $\vt^a_r$ is a complex, positive chirality SO(8) spinor and $\Pi_{ab}\equiv(\g^1\g^2\g^3\g^4)_{ab}$
is symmetric, traceless and squares to one.
The mode expansions of $\vt^a_r$ and its conjugate momentum $\l^a_r$ at $\t=0$ are
\begin{equation}
\begin{split}
\vt^a_r(\s_r) & =\vt^a_{0(r)}+\sqrt{2}\sum_{n=1}^{\infty}
\bigl(\vt^a_{n(r)}\cos\frac{n\s_r}{|\a_r|}+\vt^a_{-n(r)}\sin\frac{n\s_r}{|\a_r|}\bigr)\,,\\
\l^a_r(\s_r) & =\frac{1}{2\pi|\a_r|}\bigl[\l^a_{0(r)}+\sqrt{2}\sum_{n=1}^{\infty}
\bigl(\l^a_{n(r)}\cos\frac{n\s_r}{|\a_r|}+\l^a_{-n(r)}\sin\frac{n\s_r}{|\a_r|}\bigr)\bigr]\,.
\end{split}
\end{equation}
The Fourier modes satisfy $2\l_{n(r)}^a=|\a_r|\bar{\vt}^a_{n(r)}$
and the canonical anti-commutation relations for the fermionic coordinates yield the anti-commutation rules
\begin{equation}
\{\vt^a_r(\s_r),\l^b_s(\s_s)\}=\d^{ab}\d_{rs}\d(\s_r-\s_s)\qquad\lr\qquad\{\vt^a_{n(r)},\l^b_{m(s)}\}=\d^{ab}\d_{nm}\d_{rs}\,.
\end{equation}
The fermionic normal modes are defined via~\cite{Spradlin:2002ar}
\begin{equation}
\vt_{n(r)}=\frac{c_{n(r)}}{\sqrt{|\a_r|}}\left[(1+\r_{n(r)}\Pi)b_{n(r)}
+e(\a_r)e(n)(1-\r_{n(r)}\Pi)b_{-n(r)}^{\dag}\right]\,,\qquad n\in\Zop\,,
\end{equation}
and break the $SO(8)$ symmetry to $SO(4)\times SO(4)$. Here
\begin{equation}
\r_{n(r)}=\r_{-n(r)}=\frac{\o_{n(r)}-|n|}{\m\a_r}\,,\qquad
c_{n(r)}=c_{-n(r)}=\frac{1}{\sqrt{1+\r_{n(r)}^2}}\,.
\end{equation}
These modes satisfy
\begin{equation}
\{b^a_{n(r)},b^{b\,\dag}_{m(s)}\}=\d^{ab}\d_{nm}\d_{rs}\,.
\end{equation}
In what follows it will be important to use a $\g$-matrix representation for which the
$SO(4)\times SO(4)$ symmetry
of the plane wave space-time is manifest for the fermions. In this basis
\begin{equation}
\Pi =
\begin{pmatrix}
\d_{\a_1}^{\b_1}\d_{\a_2}^{\b_2} & 0 \\ 0 & -\d^{\da_1}_{\db_1}\d^{\da_2}_{\db_2}
\end{pmatrix}\,,
\end{equation}
where $\a_k$, $\da_k$ ($\b_k$, $\db_k$) are two-component Weyl indices of $SO(4)_k$, see appendix~\ref{appA} for our conventions.
Hence $(1\pm\Pi)/2$
project onto $({\bf 2},{\bf 2})$ and $({\bf 2'},{\bf 2'})$ of $SO(4)\times SO(4)$,
respectively, and
\begin{equation}
\{b_{n(r)\,\a_1\a_2},b^{\b_1\b_2\,\dag}_{m(s)}\}=\d_{\a_1}^{\b_1}\d_{\a_2}^{\b_2}\d_{nm}\d_{rs}\,,\qquad
\{b_{n(r)\,\da_1\da_2},b^{\db_1\db_2\,\dag}_{m(s)}\}=\d_{\da_1}^{\db_1}\d_{\da_2}^{\db_2}\d_{nm}\d_{rs}\,.
\end{equation}
The free string light-cone Hamiltonian is
\begin{equation}
H_{2(r)}=\frac{1}{\a_r}\sum_{n\in\Zop}\o_{n(r)}\bigl(a_{n(r)}^{I\,\dag}a_{n(r)}^I
+b^{\a_1\a_2\,\dag}_{n(r)}b_{n(r)\,\a_1\a_2}+b^{\da_1\da_2\,\dag}_{n(r)}b_{n(r)\,\da_1\da_2}\bigr)\,.
\end{equation}
In the above the zero-point energies cancel between bosons and fermions. The vacuum $|v\ra_r$ is defined as
\begin{equation}
a_{n(r)}|v\ra_r=0\,,\qquad b_{n(r)}|v\ra_r=0\,,\quad n\in\Nop\,.
\end{equation}
We will sometimes suppress the spinor indices and go back and forth between expressions which are formally $SO(8)$ invariant but
contain $\Pi$ and those where the $SO(4)\times SO(4)$ symmetry is manifest.

The isometries of the plane wave superalgebra are generated by $H$, $P^+$, $J^{+I}$,
$J^{ij}$ and $J^{i'j'}$. The latter two are angular momentum generators of the transverse $SO(4)\times SO(4)$ symmetry
of the plane wave space-time. The 32 supersymmetries are generated by $Q^+$, $\bar{Q}^+$ and $Q^-$, $\bar{Q}^-$.
The former correspond to inhomogeneous shift symmetries on the world-sheet (`non-linearly realized' supersymmetries), whereas the
latter generate the linearly realized world-sheet supersymmetries.
An important subset of the superalgebra is~\cite{Blau:2001ne}
\begin{equation}\label{comm}
\begin{split}
[H,P^I] &= -i\m^2J^{+I}\,,\qquad [H,Q^+]=-\m\Pi Q^+\,,\\
\{Q^-_{\da},\bar{Q}^-_{\db}\} &=
2\d_{\dot{a}\dot{b}}H-i\m\bigl(\g_{ij}\Pi\bigr)_{\dot{a}\dot{b}}J^{ij}+i\m\bigl(\g_{i'j'}\Pi\bigr)_{\dot{a}\dot{b}}J^{i'j'}\,.
\end{split}
\end{equation}
The supercharges, obtained by the standard Noether method in~\cite{Metsaev:2001bj}, are
\begin{align}
\label{Q+}
Q^+_{(r)} & = \sqrt{\frac{2}{\a'}}\int_0^{2\pi|\a_r|}d\s_r\,\sqrt{2}\l_r\,,\\
\label{qfield}
Q^-_{(r)} & =\sqrt{\frac{2}{\a'}}\int_0^{2\pi|\a_r|}d\s_r\,\left[2\pi\a'e(\a_r)p_r\g\l_r-ix'_r\g\bar{\l}_r-i\m x_r\g\Pi\l_r\right]\,,
\end{align}
and $\bar{Q}^{\pm}_{(r)}=e(\a_r)\bigl[Q_{(r)}^{\pm}\bigr]^{\dag}$.
Expanding $Q^-$ in modes one finds
\begin{equation}\label{q-mode}
\begin{split}
Q^-_{(r)} & =\frac{e(\a_r)}{\sqrt{|\a_r|}}\g
\Bigl(\sqrt{\m}\left[a_{0(r)}(1+e(\a_r)\Pi)+a_{0(r)}^{\dag}(1-e(\a_r)\Pi)\right]\l_{0(r)}\\
&+\sum_{n\neq 0}\sqrt{|n|}\left[a_{n(r)}P_{n(r)}^{-1}b_{n(r)}^{\dag}
+e(\a_r)e(n)a_{n(r)}^{\dag}P_{n(r)}b_{-n(r)}\right]\Bigr)\,,
\end{split}
\end{equation}
where
\begin{equation}
P_{n(r)}\equiv\frac{1-\r_{n(r)}\Pi}{\sqrt{1-\r_{n(r)}^2}}
=\frac{1+\Pi}{2}U_{|n|(r)}^{1/2}+\frac{1-\Pi}{2}U_{|n|(r)}^{-1/2}\,,\qquad
U_{n(r)}\equiv\frac{\o_{n(r)}-\m\a_r}{n}\,.
\end{equation}

\subsection{The kinematical part of the vertex}

The guiding principles in the construction of light-cone string field theory are world-sheet
continuity and the realization of the superalgebra in the full interacting theory. One can distinguish two
sets of generators. One consists of the kinematical generators
$P^+$, $P^I$, $J^{+I}$, $J^{ij}$, $J^{i'j'}$, $Q^+$ and $\bar{Q}^+$
which are not corrected by interactions, in other words the symmetries they generate are not
affected by adding higher order terms to the action. On the other hand, the dynamical generators
$H$, $Q^-$ and $\bar{Q}^-$
do receive corrections in the presence of interactions. The requirement that the
superalgebra is satisfied in the interacting theory, now gives rise to two kinds of constraints: {\em kinematical} constraints arising
from the (anti)commutation relations of kinematical with dynamical generators and {\em dynamical} constraints arising from the
(anti)commutation relations of dynamical generators alone. The former lead to the continuity conditions
in superspace, whereas the latter require the insertion of interaction point operators~\cite{Mandelstam:1974hk,Green:1983tc}.
In practice these constraints will be
solved in perturbation theory, for example $H$, the full Hamiltonian of the interacting theory, has an expansion in the string coupling
\begin{equation}\label{hexp}
H=H_2+g_{\text{s}}H_3+\cdots\,,
\end{equation}
and $H_3$ leads to a three-string interaction. Then it follows that e.g.
\begin{equation}
[H,P^I] = -i\m^2J^{+I}\quad\Longrightarrow\quad[H_3,P^I]=0\,,
\end{equation}
so the interaction vertex is translationally invariant and conserves transverse momentum (which is not a good quantum number in the plane
wave space-time due to the confining harmonic oscillator potential). This constraint is the same as
world-sheet continuity in momentum space and will be realized by constructing an interaction vertex which
conserves $\sum_rp_r(\s_r)$ locally~\cite{Green:1983tc}. Analogously the fermionic world-sheet continuity condition follows from
\begin{equation}\label{q+}
[H,Q^+]=-\m\Pi Q^+\quad\Longrightarrow\quad[H_3,Q^+]=0\,,
\end{equation}
and will be implemented by local conservation of $\sum_r\l_r(\s_r)$, cf.~\eqref{Q+}.
In practice it is useful to express e.g.\ $H_3$ not as an operator, but as a state
$|H_3\ra$ in the multi-string Hilbert space and work in the number basis~\cite{Cremmer:1974jq,Cremmer:1975ej}. Then the dynamical
generators are of the form ${\mc P}|V\ra$,
where ${\mc P}$ are the prefactors determined by the dynamical constraints (i.e. the oscillator expressions of the interaction point
operators mentioned above) and the kinematical part of the vertex $|V\ra$ common to all
the dynamical generators implements the continuity conditions. In the number basis it has the form
\begin{equation}
|V\ra\equiv|E_a\ra|E_b\ra\d\bigl(\sum_{r=1}^3\a_r\bigr)\,,
\end{equation}
where $|E_a\ra$ and $|E_b\ra$ are exponentials of bosonic and fermionic creation oscillators,
respectively.
The bosonic solution to the kinematic constraints is~\cite{Spradlin:2002ar}
\begin{equation}\label{bv}
|E_a\ra = \exp\left(\frac{1}{2}
\sum_{r,s=1}^3\sum_{m,n\in\Zop}a^{I\,\dag}_{m(r)}\bar{N}^{rs}_{mn}a^{I\,\dag}_{n(s)}
\right)|v\ra_{123}\,,
\end{equation}
where $|v\ra_{123}=|v\ra_1\otimes|v\ra_2\otimes|v\ra_3$ is the tensor product of three (bosonic) vacuum states and $\bar{N}^{rs}_{mn}$
are the so called Neumann matrices, see e.g.~\cite{Spradlin:2002ar,Pankiewicz:2002tg} for their formal definitions and~\cite{He:2002zu}
for explicit expressions as functions of $\m$, $\a_r$.
A solution of the fermionic kinematical constraints built on $|v\ra_{123}$
(instead of $|0\ra_{123}$~\cite{Spradlin:2002ar,Pankiewicz:2002gs}, cf.~the discussion in the introduction)
is~\cite{Chu:2002eu,Chu:2002wj}\,\footnote{To compare with the expression in~\cite{Chu:2002wj}, note that
\begin{equation}
\sum_{s,t}\e^{st}\sqrt{\a_s}\a_tb_{0(s)}^{\dag\,\a_1\a_2} =-\sqrt{2}\L^{\a_1\a_2}\,,\qquad
\sum_{s,t}\e^{st}\sqrt{\a_s}\a_tb_{0(s)}^{\dag\,\da_1\da_2}=\frac{\a}{\sqrt{2}}\Theta^{\da_1\da_2}\,.
\end{equation}}
\begin{equation}\label{eb2}
|E_b\ra=\exp\left(
\sum_{r,s=1}^3\sum_{m,n\ge0}
\bigl(b^{\a_1\a_2\,\dag}_{-m(r)}b^{\dag}_{n(s)\,\a_1\a_2}+b^{\da_1\da_2\,\dag}_{m(r)}b^{\dag}_{-n(s)\,\da_1\da_2}\bigr)
\bar{Q}^{rs}_{mn}\right)|v\ra_{123}\,,
\end{equation}
where the non-vanishing components of the fermionic Neumann matrices are~\cite{Chu:2002eu,Pankiewicz:2002gs} ($m$, $n>0$; no sum
over $r$ in~\eqref{q2} and we use the notations of~\cite{Pankiewicz:2002gs})
\begin{align}
\label{q1}
\bar{Q}^{rs}_{mn} & =
e(\a_r)\sqrt{\left|\frac{\a_s}{\a_r}\right|}\bigl[U_{(r)}^{1/2}C^{1/2}\bar{N}^{rs}C^{-1/2}U_{(s)}^{1/2}\bigr]_{mn}\,,\\
\label{q2}
\bar{Q}^{rs}_{0n} & = \e^{rt}\sqrt{\a_r}\a_t\frac{e(\a_s)}{\sqrt{|\a_s|}}\bigl[\bigl(U_{(s)}C_{(s)}C\bigr)^{1/2}\bar{N}^s\bigr]_n\,,\\
\label{q3}
\bar{Q}^{3r}_{00} & = -\bar{Q}^{r3}_{00} = \frac{1}{2}\sqrt{-\frac{\a_r}{\a_3}}\,.
\end{align}

\section{A supersymmetric extension in the $SO(4)\times SO(4)$ \\ formalism}\label{section3}

\subsection{The superalgebra and the constituents of the prefactor}

It is convenient to define the linear combinations of the free supercharges ($\eta=e^{i\pi/4}$)
\begin{equation}
\sqrt{2}\eta\,Q\equiv Q^-+i\bar{Q}^-\,,\qquad\text{and}\qquad
\sqrt{2}\bar{\eta}\,\wt{Q}=Q^--i\bar{Q}^-
\end{equation}
which, on the subspace of physical states satisfying the level-matching
condition~\cite{Green:1983tc} (see also the recent~\cite{DiVecchia:2003yp}), satisfy
\begin{equation} 
\begin{split}
\{Q_{\dot{a}},Q_{\dot{b}}\}& = \{\wt{Q}_{\dot{a}},\wt{Q}_{\dot{b}}\}
=2\d_{\dot{a}\dot{b}}H\,,\\
\{Q_{\dot{a}},\wt{Q}_{\dot{b}}\} & =
-\m\bigl(\g_{ij}\Pi\bigr)_{\dot{a}\dot{b}}J^{ij}
+\m\bigl(\g_{i'j'}\Pi\bigr)_{\dot{a}\dot{b}}J^{i'j'}\,.
\end{split}
\end{equation}
Since $J^{ij}$ and $J^{i'j'}$ are not corrected by the interaction,
it follows that at order ${\mc O}(g_{\text{s}})$ the dynamical generators have to obey
\begin{align}
\label{dyn1}
\sum_{r=1}^3Q_{\dot{a}(r)}|Q_{3\,\dot{b}}\ra+\sum_{r=1}^3Q_{\dot{b}(r)}|Q_{3\,\dot{a}}\ra
& = 2|H_3\ra\d_{\dot{a}\dot{b}}\,,\\
\label{dyn2}
\sum_{r=1}^3\wt{Q}_{\dot{a}(r)}|\wt{Q}_{3\,\dot{b}}\ra+\sum_{r=1}^3\wt{Q}_{\dot{b}(r)}|\wt{Q}_{3\,\dot{a}}\ra
&= 2|H_3\ra\d_{\dot{a}\dot{b}}\,,\\
\label{dyn3}
\sum_{r=1}^3Q_{\dot{a}(r)}|\wt{Q}_{3\,\dot{b}}\ra+\sum_{r=1}^3\wt{Q}_{\dot{b}(r)}|Q_{3\,\dot{a}}\ra  &=0\,.
\end{align}
Rewriting these constraints in $SO(4)\times SO(4)$ notation,
each gives rise to four constraints, e.g.~\eqref{dyn1} yields
\begin{align}
\label{1}
\sum_rQ_{(r)\,\a_1\da_2}|Q_{3\,\b_1\db_2}\ra+\sum_r Q_{(r)\,\b_1\db_2}|Q_{3\,\a_1\da_2}\ra &
= -2\e_{\a_1\b_1}\e_{\da_2\db_2}|H_3\ra\,,\\
\label{2}
\sum_rQ_{(r)\,\da_1\a_2}|Q_{3\,\db_1\b_2}\ra+\sum_r Q_{(r)\,\db_1\b_2}|Q_{3\,\da_1\a_2}\ra & = -2\e_{\da_1\db_1}\e_{\a_2\b_2}|H_3\ra\,,\\
\label{3}
\sum_rQ_{(r)\,\a_1\da_2}|Q_{3\,\db_1\b_2}\ra+\sum_r Q_{(r)\,\db_1\b_2}|Q_{3\,\a_1\da_2}\ra & = 0\,,
\end{align}
with the fourth constraint equivalent to the third one.
So as not to destroy world-sheet continuity, the prefactors have to (anti)commute with the kinematical constraints.
The relevant bosonic combinations that are linear in creation oscillators are
\begin{equation}
K^I \equiv \sum_{r=1}^3\sum_{n\in\Zop}K_{n(r)}a_{n(r)}^{I\,\dag}\,,\qquad
\K^I \equiv \sum_{r=1}^3\sum_{n\in\Zop}\K_{n(r)}a_{n(r)}^{I\,\dag}\,,
\end{equation}
where
\begin{equation}
K_{n(r)} \equiv (1-4\m\a K)^{1/2}\begin{cases} F_{0(r)}\,, & n=0 \\
F_{n(r)}\,, & n>0 \\
iU_{n(r)}F_{n(r)}\,, & n<0
\end{cases}
\end{equation}
and $\K_{n(r)}=K^*_{n(r)}$. For the explicit expressions of $K$ and $F_{n(r)}$ see e.g.\ \cite{Spradlin:2002rv,Pankiewicz:2002gs}.
The fermionic expressions anticommuting with the kinematical constraints that are linear in creation oscillators are
(see also~\cite{DiVecchia:2003yp})\,\footnote{There exist further expressions anticommuting with the kinematical
constraints~\cite{DiVecchia:2003yp}, which however can be written as $\p_{\s}Y$ and $\p_{\s}Z$, cf. appendices~\ref{appC} and~\ref{appD}.}
\begin{equation}
\label{Y}
Y^{\a_1\a_2} = \sum_{r=1}^3\sum_{n\ge 0}\bar{G}_{n(r)}b_{n(r)}^{\dag\,\a_1\a_2}\,,\qquad
Z^{\da_1\da_2} = \sum_{r=1}^3\sum_{n\ge 0}\bar{G}_{n(r)}b_{-n(r)}^{\dag\,\da_1\da_2}\,,
\end{equation}
where
\begin{equation}\label{G}
\bar{G}_{(r)} =
\sqrt{-\frac{\a'}{\a}}(1-4\m\a K)^{1/2}\sqrt{|\a_r|}U_{(r)}^{1/2}C^{-1/2}F_{(r)}\,,
\end{equation}
and we have chosen a normalization convenient for what follows.
In order to derive equations that determine the dynamical
generators one has to compute (anti)commutators of $Q_{\dot{a}(r)}$ and $\wt{Q}_{\dot{a}(r)}$
with the constituents of the prefactors. Moreover, the action of the supercharges on
$|V\ra$ has to be known in terms of these constituents.
To determine the equations arising from the constraints~\eqref{dyn1} and~\eqref{dyn2} we
will need
\begin{align}
\label{qtk}
& \eta\sqrt{\frac{-2\a}{\a'}}\sum_{r=1}^3[Q_{(r)},\wt{K}^I]=
-2\frac{\m\a}{\a'}\g^I\left(Y+iZ\right)\,,\\
\label{tqk}
& \bar{\eta}\sqrt{\frac{-2\a}{\a'}}\sum_{r=1}^3[\wt{Q}_{(r)},K^I]=
-2\frac{\m\a}{\a'}\g^I\left(Y-iZ\right)\,.
\end{align}
Furthermore
\begin{equation}\label{qy}
\begin{split}
\eta\sqrt{\frac{-2\a}{\a'}}\sum_{r=1}^3\{Q_{(r)},Y+iZ\} & = i\g^IK^I\,,\\
\bar{\eta}\sqrt{\frac{-2\a}{\a'}}\sum_{r=1}^3\{\wt{Q}_{(r)},Y-iZ\} & = -i\g^I\wt{K}^I\,.
\end{split}
\end{equation}
Finally, the action of the supercharges on $|V\ra$ is
\begin{align}
\label{qv}
\eta\sqrt{\frac{-2\a}{\a'}}\sum_{r=1}^3Q_{(r)}|V\ra & = K^I\g^I(Y+iZ)|V\ra\,,\\
\label{tqv}
\bar{\eta}\sqrt{\frac{-2\a}{\a'}}\sum_{r=1}^3\wt{Q}_{(r)}|V\ra & =\K^I\g^I(Y-iZ)|V\ra\,.
\end{align}
These equations can be written in $SO(4)\times SO(4)$ notation, the resulting formulae are relegated to appendix~\ref{appB}.
It is useful to note the reality properties under complex conjugation defined as
\begin{equation}
{K^*}^I\equiv\sum_{r=1}^3\sum_{n\in\Zop}K^*_{n(r)}a_{n(r)}^{I\,\dag} = \K^I,\quad
Y^*\equiv\sum_{r=1}^3\sum_{n\ge 0}\bar{G}^*_{n(r)}b_{n(r)}^{\dag}=Y,\quad
Z^*\equiv\sum_{r=1}^3\sum_{n\ge 0}\bar{G}^*_{n(r)}b_{-n(r)}^{\dag} = Z,
\end{equation}
which enable us to immediately write down the solution for $|\wt{Q}\ra$, once we
determined $|Q\ra$.

\subsection{The dynamical generators at ${\mc O}(g_{\text{s}})$}

To construct a solution to the dynamical constraints we will proceed as follows: first
we write down an ansatz for, say $|Q_{3\,\b_1\db_2}\ra$, which is of the form
\begin{equation}\label{ansatz1}
|Q_{3\,\b_1\db_2}\ra =
\bigl(f^i_{\b_1\db_2}(Y,Z)\K^i+g^{i'}_{\b_1\db_2}(Y,Z)\K^{i'}\bigr)|V\ra\,,
\end{equation}
where $f^i_{\b_1\db_2}(Y,Z)$ and $g^{i'}_{\b_1\db_2}(Y,Z)$ are the most general polynomials in
$Y$ and $Z$ compatible with the index structure. Computing the left-hand-side of
equation~\eqref{1} and requiring that the result only
contains the tensor $\e_{\a_1\b_1}\e_{\da_2\db_2}$ fixes all the coefficients
except for the overall normalization and analogously for~\eqref{2};
the relative normalization between
$|Q_{3\,\b_1\db_2}\ra$ and $|Q_{3\,\db_1\b_2}\ra$ is fixed by demanding that the
right-hand-sides of~\eqref{1} and~\eqref{2} result in the same cubic interaction vertex.
Finally, one can check that~\eqref{3} is satisfied with these fixed coefficients.

To determine the most general ansatz for the fermionic polynomials we define the following
quantities which are quadratic in $Y$ and symmetric in spinor indices
\begin{equation}\label{y2}
Y^2_{\a_1\b_1} \equiv Y_{\a_1\a_2}Y^{\a_2}_{\b_1}\,,\qquad Y^2_{\a_2\b_2} \equiv Y_{\a_1\a_2}Y^{\a_1}_{\b_2}\,,
\end{equation}
cubic in $Y$
\begin{equation}\label{y3}
Y^3_{\a_1\b_2} \equiv Y^2_{\a_1\b_1}Y^{\b_1}_{\b_2}=-Y^2_{\b_2\a_2}Y^{\a_2}_{\a_1}\,,
\end{equation}
and, finally, quartic in $Y$ and antisymmetric in spinor indices
\begin{equation}
Y^4_{\a_1\b_1} \equiv Y^2_{\a_1\g_1}{Y^2}^{\g_1}_{\b_1}=-\frac{1}{2}\e_{\a_1\b_1}Y^4\,,\qquad
Y^4_{\a_2\b_2} \equiv Y^2_{\a_2\g_2}{Y^2}^{\g_2}_{\b_2}=\frac{1}{2}\e_{\a_2\b_2}Y^4\,,
\end{equation}
where
\begin{equation}\label{y4}
Y^4 \equiv Y^2_{\a_1\b_1}{Y^2}^{\a_1\b_1}=-Y^2_{\a_2\b_2}{Y^2}^{\a_2\b_2}\,.
\end{equation}
These quantities satisfy various useful relations given in appendix B,~\eqref{id1}--\eqref{id4}.
Analogous definitions are made for $Z$.
Performing the above procedure (cf.\ appendices~\ref{appB} and~\ref{appC} for details) we
find the following solution for the dynamical supercharges
\begin{equation}
\bar{\eta}\sqrt{\frac{\a'}{-2\a}}|Q_{3\,\b_1\db_2}\ra =
\Bigl(s_{\dg_1\db_2}(Z)t_{\b_1\g_1}(Y)\K^{\dg_1\g_1}+is_{\b_1\g_2}(Y)t^*_{\db_2\dg_2}(Z)\K^{\dg_2\g_2}\Bigr)|V\ra\,,
\end{equation}
\begin{equation}
-\eta\sqrt{\frac{\a'}{-2\a}}|Q_{3\,\db_1\b_2}\ra =
\Bigl(s^*_{\g_1\b_2}(Y)t^*_{\db_1\dg_1}(Z)\K^{\dg_1\g_1}+is^*_{\db_1\dg_2}(Z)t_{\b_2\g_2}(Y)\K^{\dg_2\g_2}\Bigr)|V\ra\,.
\end{equation}
Here we defined
\begin{equation}
\K^{\dg_1\g_1} \equiv \K^i{\s^i}^{\dg_1\g_1}\,,\qquad
\K^{\dg_2\g_2} \equiv \K^{i'}{\s^{i'}}^{\dg_2\g_2}\,,
\end{equation}
and the spinorial quantities, say for $Y$
\begin{equation}
s(Y) \equiv Y+\frac{i}{3}Y^3\,,\qquad t(Y) \equiv \e+iY^2-\frac{1}{6}Y^4\,.
\end{equation}
The remaining dynamical supercharges are $|\wt{Q}\ra=|Q^*\ra$, some details for the proof
of the remaining constraint~\eqref{dyn3} are given in appendix~\ref{appC}.
The cubic interaction vertex is
\begin{align}
|H_3\ra & =
\Bigl[\bigl(K_i\K_j-\frac{\m\a}{\a'}\d_{ij}\bigr)v^{ij}
-\bigl(K_{i'}\K_{j'}-\frac{\m\a}{\a'}\d_{i'j'}\bigr)v^{i'j'}
\nonumber\\
&-K^{\da_1\a_1}\K^{\da_2\a_2}s_{\a_1\a_2}(Y)s^*_{\da_1\da_2}(Z)
-\K^{\da_1\a_1}K^{\da_2\a_2}s^*_{\a_1\a_2}(Y)s_{\da_1\da_2}(Z)\Bigr]|V\ra\,,
\end{align}
where\footnote{The first line can also be written as 
\begin{align}
\bigl(K_i\K_j-\frac{\m\a}{\a'}\d_{ij}\bigr)v^{ij}-\bigl(K_{i'}\K_{j'}-\frac{\m\a}{\a'}\d_{i'j'}\bigr)v^{i'j'} & =
\bigl(\frac{1}{2}K^{\da_1\a_1}\K^{\db_1\b_1}-\frac{\m\a}{\a'}\e^{\a_1\b_1}\e^{\da_1\db_1}\bigr)t_{\a_1\b_1}(Y)t^*_{\da_1\db_1}(Z)\nonumber\\
&-\bigl(\frac{1}{2}K^{\da_2\a_2}\K^{\db_2\b_2}-\frac{\m\a}{\a'}\e^{\a_2\b_2}\e^{\da_2\db_2}\bigr)t_{\a_2\b_2}(Y)t^*_{\da_2\db_2}(Z)\,.
\end{align}}
\begin{align}
v^{ij} & =
\d^{ij}\Bigl[1+\frac{1}{12}\bigl(Y^4+Z^4\bigr)+\frac{1}{144}Y^4Z^4\Bigr]\nonumber\\
&-\frac{i}{2}\Bigl[{Y^2}^{ij}\bigl(1+\frac{1}{12}Z^4\bigr)
-{Z^2}^{ij}\bigl(1+\frac{1}{12}Y^4\bigr)\Bigr]
+\frac{1}{4}\bigl[Y^2Z^2\bigr]^{ij}\,,\\
v^{i'j'} & =
\d^{i'j'}\Bigl[1-\frac{1}{12}\bigl(Y^4+Z^4\bigr)+\frac{1}{144}Y^4Z^4\Bigr]\nonumber\\
&-\frac{i}{2}\Bigl[{Y^2}^{i'j'}\bigl(1-\frac{1}{12}Z^4\bigr)
-{Z^2}^{i'j'}\bigl(1-\frac{1}{12}Y^4\bigr)\Bigr]
+\frac{1}{4}\bigl[Y^2Z^2\bigr]^{i'j'}\,.
\end{align}
Here we defined
\begin{equation}
{Y^2}^{ij} \equiv \s^{ij}_{\a_1\b_1}{Y^2}^{\a_1\b_1}\,,\quad
{Z^2}^{ij} \equiv \s^{ij}_{\da_1\db_1}{Z^2}^{\da_1\db_1}\,,\quad
\bigl(Y^2Z^2\bigr)^{ij} \equiv {Y^2}^{k(i}{Z^2}^{j)k}
\end{equation}
and analogously for the primed indices. Notice that
$v_{ij}^* = v_{ji}$, $v_{i'j'}^* = v_{j'i'}$ and
$\d_{ij}v^{ij}-\d_{i'j'}v^{i'j'}$ only
yields a contribution quartic in $Y$ and $Z$.

From the purely bosonic part of the prefactor it is manifest
that it has {\em negative} parity under the discrete $\Zop_2$ symmetry
of the plane wave space-time. Indeed, taking the overall normalization (which cannot be fixed by the dynamical constraints) to be
$-\pi g_{\text{s}}\a'\m^2$, the effective interaction vertex for
states containing only bosonic oscillators is
\begin{align}\label{heff}
&-\pi g_{\text{s}}\a'\m^2|H_{3\,\text{bos}}\ra =
-\pi g_{\text{s}}\a'\m^2\bigl(K^i\K^i-K^{i'}\K^{i'}\bigr)|E_a\ra\d\bigl(\sum_{r=1}^3\a_r\bigr)
\nonumber\\
&=\frac{1}{2}g_2\b(\b+1)\sum_{r=1}^3\sum_{n\in\Zop}\frac{\o_{n(r)}}{\a_r}
\bigl(a^{i\,\dag}_{n(r)}a^i_{-n(r)}-a^{i'\,\dag}_{n(r)}a^{i'}_{-n(r)}\bigr)|E_a\ra|\a_3|
\d\bigl(\sum_{r=1}^3\a_r\bigr)\,.
\end{align}
In the second equality we have used an identity derived
in~\cite{Pearson:2002zs,Lee:2002vz}. The effective vertex is {\em identical}
to that obtained in the $SO(8)$ formalism (there one has to take into account an
expectation value of fermionic zero-modes, see e.g.~\cite{Pankiewicz:2002tg}).
In particular it follows that we have to assign negative $\Zop_2$-parity to the vacuum states $|v\ra_r$
if the solution we have found should preserve all the transverse symmetries of the plane wave. Then, the vertex in the $SO(8)$
formalism preserves all the symmetries as well, since the parity of $|0\ra$ is positive. In~\cite{Chu:2002eu} it was
proposed that the parity of the plane wave vacuum $|v\ra$ should be positive in order to obtain an interaction vertex
in the $SO(4)\times SO(4)$ formalism that preserves
the $\Zop_2$ symmetry. Then the vertex of~\cite{Spradlin:2002ar,Spradlin:2002rv,Pankiewicz:2002gs,Pankiewicz:2002tg}
would break the $\Zop_2$-symmetry.
This proposal however, was based on the assertion that the constant part of the fermionic
prefactor is
$SO(8)$ invariant. Now that we have solved the constraints we see that this is not the case and, therefore, the full transverse
symmetry is preserved in {\em both} formulations.
Further evidence for the above assignment of parity can be extracted from earlier literature:
the spectrum of type IIB string theory on the plane wave was analyzed in detail
in~\cite{Metsaev:2002re}, in particular the precise correspondence between the lowest lying string
states and the fluctuation modes of supergravity on the plane wave was established.
As mentioned in the introduction, the state $|0\ra$ corresponds to a complex scalar arising from the
dilaton-axion system, whereas the state $|v\ra$ corresponds to a complex scalar
being a mixture of the trace of the graviton and the \RR potential
on one of the $\Rop^4$'s, i.e. the chiral primary sector. Since the dilaton and axion are scalars under $SO(8)$ and
the discrete $\Zop_2$ is just a particular $SO(8)$ transformation, the
assignment of positive parity to $|0\ra$ appears to be the correct one. Moreover, analysis
of the interaction Hamiltonian for the chiral primary sector shows that invariance of the
Hamiltonian under the $\Zop_2$ requires the chiral primaries to have negative
parity~\cite{Kiem:2002pb}.

It is therefore natural to conjecture that the two apparently different formulations for light-cone string field theory
are physically equivalent. In the next section we will present additional evidence for this claim
by computing the leading order mass-shift of the simplest stringy state
and showing that at leading order in large $\m$ the torus anomalous dimension of the dual
gauge theory operators is again reproduced (as in the $SO(8)$ formalism~\cite{Roiban:2002xr,Gomis:2003kj}).

We know that the vertex of~\cite{Spradlin:2002ar,Spradlin:2002rv,Pankiewicz:2002gs,Pankiewicz:2002tg} reduces to
the one in flat space when $\m\to 0$.
It is quite plausible that even in
flat space one could construct a solution to the constraints starting with {\em any} state in the
massless multiplet, e.g.\ in~\cite{Green:1984fu} the closed string cubic interaction vertex
was constructed in the $SU(4)$ formalism, where only a $SU(4)\times U(1)$ subgroup of the
transverse $SO(8)$ symmetry is manifest.
In flat space the most natural approach is the $SO(8)$ formalism,
since there the maximal transverse symmetry is manifest. In the plane wave however, there are two natural choices: to have
a manifest flat space limit (and therefore to be able to work with quantities that are
{\em formally} $SO(8)$ invariant) or to have the true transverse symmetry manifest, but give up
the (manifest) smooth connection to the flat space vertex of the $SO(8)$ formalism; both approaches have their advantages,
e.g., as we will see in the next section, it is simpler to compute
correlators with fermionic oscillators in the formalism presented here.

Note that our reasoning implies that if the plane wave ground state
$|v\ra$ is odd under the $\Zop_2$ then the recent solution
to the dynamical constraints given in~\cite{DiVecchia:2003yp}, 
which is of the form $|H_3\ra\sim\p_{\t}|V\ra$, is only $SO(4)\times SO(4)$ but not
$\Zop_2$ invariant. If, on the contrary, one insists on assigning positive $\Zop_2$ parity to $|v\ra$~\cite{Chu:2002eu}, 
then the $\Zop_2$
invariant vertex would be given in~\cite{DiVecchia:2003yp}.
It would be nice to understand the origin for the successful tests of the vertex-correlator duality reported
in~\cite{Chu:2003qd,Georgiou:2003aa,Chu:2003ji} given the fact that the `phenomenological' prefactor proposed in~\cite{Chu:2003qd}
does not agree with the result found here (nor with the one of~\cite{DiVecchia:2003yp}).

\section{Anomalous dimension from string field theory}\label{section4}

In this section we compute the mass-shift due to interactions of the simplest stringy state
\footnote{The
change of basis $\a_n=\frac{1}{\sqrt{2}}\bigl(a_{|n|}+ie(n)a_{-|n|}\bigr)$
for $n\neq0$ is for convenience.}
\begin{equation}\label{nij}
|n\ra\equiv\a^{I\,\dag}_{n(3)}\a^{J\,\dag}_{-n(3)}|v\ra_3\,.
\end{equation}
using non-degenerate perturbation theory. In principle one should use degenerate perturbation theory as the
single string state can mix with multi-string states having the same energy.  The same caveat
holds for the computation in gauge theory and we will ignore this complication here.
In the $SO(8)$ formalism this has been done for the symmetric-traceless ${\bf 9}$ and
antisymmetric ${\bf 6}={\bf 3}+\bar{{\bf 3}}$ of
either one of the $SO(4)$'s in~\cite{Roiban:2002xr} and for the trace ${\bf 1}$
in~\cite{Gomis:2003kj}. Here we repeat this computation in the $SO(4)\times SO(4)$
formulation constructed in the previous section and also extend the
analysis to the $({\bf 4},{\bf 4})_{\pm}$\footnote{We define the states in  $({\bf 4},{\bf 4})_{\pm}$
as $\frac{1}{2}\bigl(\a_{n(3)}^{i\,\dag}\a_{-n(3)}^{j'\,\dag}\pm\a_{-n(3)}^{i\,\dag}\a_{n(3)}^{j'\,\dag}\bigr)|v\ra_3$.}
of $SO(4)\times SO(4)$. These
correspond to BMN operators with mixed scalar/vector impurities and
superconformal symmetry of the gauge theory implies that they
have the same anomalous dimension as the other representations~\cite{Beisert:2002tn}.

At lowest order the eigenvalue correction comes from two contributions; one-loop diagram and contact term
\begin{equation}\label{de1}
\d E_n^{(2)}\la n|n\ra = g_2^2\sum_{1,2}
\frac{1}{2}\frac{\left|\la n|H_3|1,2\ra\right|^2}{E_n^{(0)}-E_{1,2}^{(0)}}
-\frac{g_2^2}{4}\la n|Q_{3\,\b_1\db_1}Q_3^{\b_1\db_1}|n\ra\,.
\end{equation}
Factors different from $g_2$ in the normalization (cf.~\eqref{heff}) are absorbed in the
definition of $H_3$ and $Q_{3\,\b_1\db_1}$, the extra factor of $1/2$ in the first term is due to the reflection symmetry of
the one-loop diagram.  The sum
over $1$,~$2$ is over physical double string states, i.e.\ those obeying the
level-matching condition and for the case at hand
$Q_3^2$ is the only relevant contribution to the quartic coupling.
As the generators are hermitian we take the absolute value squared of the matrix elements
(also for the contact term after inserting a suitable projection operator).
Time-reversal in the plane wave background consists in
the transformation
\begin{equation}\label{tr}
x^+ \to -x^+\,,\qquad x^- \to - x^-\,,\qquad \m \to -\m\,,
\end{equation}
in particular the reversal of $\m$ is needed due to the presence of the \RR flux.
Up to now we have always assumed
that $\m$ is non-negative and $\a_3<0$, $\a_1$, $\a_2>0$. This is, say, the process where a
single string {\em splits} into two strings.
One can show that for the process in which
two strings {\em join} to form a single string, i.e.\ $\a_1$, $\a_2<0$ and $\a_3>0$, one should make the additional replacements
\begin{equation}
\m \to -\m\,,\qquad \Pi \to -\Pi
\end{equation}
in all expressions.
This is in agreement with
equation~\eqref{tr}. Notice that the
transformation of $\Pi$ is needed to leave the fermionic mass term invariant.
From the formal expressions for the Neumann matrices it is not manifest that the cubic corrections to the dynamical generators are
hermitian as they have to be. However, from the explicit expressions for the Neumann
matrices~\cite{He:2002zu} one can see that
all the quantities are in fact invariant under the time-reversal.
The string states obey the delta-function normalization
$\la n|n'\ra = {\mc N}|\a_3|\d(\a_3-\a_4)$, where
${\mc N}=\frac{1}{2}(1+\d^{ij})$ for the ${\bf 9}$, ${\mc N}=\frac{1}{4}\d^{ij}$ for the ${\bf 1}$
and ${\mc N}=\frac{1}{2}$ otherwise.
The sum over double string states includes a double integral
over light-cone momenta, one integral is trivial due to the string-length conservation of the
cubic interaction and the factor of $|\a_3|\d(\a_3-\a_4)$ can be cancelled on both sides of
equation~\eqref{de1}.
The remaining sum is then the usual completeness relation for harmonic
oscillators projected on physical states and we have ($\b\equiv\a_1/\a_3$)
\begin{equation}\label{de2}
{\mc N}\d E_n^{(2)}=-g_2^2\int_{-1}^0\frac{d\b}{\b(\b+1)}
\sum_{\text{modes}}\left[
\frac{1}{2}\frac{\left|\la n|H_3|1,2\ra\right|^2}{E_n^{(0)}-E_{1,2}^{(0)}}
-\frac{1}{4}\la n|Q_{3\,\b_1\db_1}|1,2\ra\la 1,2|Q_3^{\b_1\db_1}|n\ra\right]\,.
\end{equation}
The measure arises due to the fact that string states are delta-function normalized.

It is important to note that in gauge theory the dilatation operator was diagonalized
within the subspace of two-impurity BMN operators in perturbation theory in
the 't Hooft coupling $\l$
and then extrapolated to $\l$, $J\to\infty$.
But it is not obvious that the large $J$ limit of the perturbation series in $\l$ has to
agree order by order with the perturbation series in $\l'$, see for
example~\cite{Klebanov:2002mp}.
Indeed there is evidence from string theory that this is not the case.
For large $\m$ the denominator of the first term in equation~\eqref{de2}
is of order ${\mc O}(\m^{-1})$ in the
impurity conserving channel, whereas it is of order ${\mc O}(\m)$ in the
impurity non-conserving one. However, as already noticed in~\cite{Spradlin:2002rv},
matrix elements where the number of impurities changes by two are of
order ${\mc O}(1)$ and, therefore potentially can contribute to the mass-shift at leading
order, that is ${\mc O}(\m g_2^2\l')$. Notice that impurity non-conserving matrix
elements being of order one, means actually ${\mc O}(\m g_2\sqrt{\l'})$ and as the overall
factor of $\m$ is simply for dimensional reasons and should not be counted when
translating to gauge theory, implies contributions
$\sim g_2\sqrt{\l'}$ to matrix elements of the dilatation operator.
It was observed in~\cite{Roiban:2002xr} that the contribution of
the impurity non-conserving channel to~\eqref{de2} is linearly divergent.
This is due to the fact that the large $\m$ limit does not commute
with the infinite sums over mode numbers;
for finite $\m$ the divergence is regularized. So a linear divergence reflects
a contribution $\sim \m g_2^2\l'(-\m\a_3)=\m g_2^2\sqrt{\l'}$ and hence of order
$g_2^2\sqrt{\l'}$ to the anomalous dimension. This constitutes a non-perturbative, `stringy'
effect.
It remains a very interesting challenge to
investigate the contribution of the impurity non-conserving channel in detail.
In principle, it is possible that besides a divergent contribution there is also a finite
one; this
would then scale as $\m g_2^2\l'$. However, to reproduce the gauge theory result
for the anomalous dimensions of two-impurity BMN operators in string theory one is led to
a truncation of equation~\eqref{de2} to the impurity conserving channel~\cite{Roiban:2002xr}.
This analysis will be performed below.

\subsection{Contribution of one-loop diagram}

The matrix element $\la n|H_3|1,2\ra$ in the impurity conserving channel
is non-zero only if the double string state contains either two bosonic or two fermionic
oscillators. The relevant projection operator is (for brevity written in $SO(8)$ form)
\begin{equation*}
\begin{split}
\a_{0(1)}^{K\,\dag}\a_{0(2)}^{L\,\dag}|v\ra\la v|\a_{0(2)}^{L}\a_{0(1)}^{K}
&+\frac{1}{2}\sum_{r,k\in\Zop}
\a_{k(r)}^{K\,\dag}\a_{-k(r)}^{L\,\dag}|v\ra\la v|\a_{-k(r)}^{L}\a_{k(r)}^{K}\\
+b_{0(1)}^{a\,\dag}b_{0(2)}^{b\,\dag}|v\ra\la v|b_{0(2)}^{b}b_{0(1)}^{a}
&+\frac{1}{2}\sum_{r,k\in\Zop}
b_{k(r)}^{a\,\dag}b_{-k(r)}^{b\,\dag}|v\ra\la v|b_{-k(r)}^{b}b_{k(r)}^{a}\,,
\end{split}
\end{equation*}
where the sum over $K$, $L$ and $a$, $b$ is understood. We get the contribution of the
first line `for free': since the effective vertex for bosonic oscillator states is the
same as in the $SO(8)$ formalism, we can use the results
of~\cite{Roiban:2002xr,Gomis:2003kj}. Using the large $\m$ expansions for the bosonic Neumann
matrices~\cite{Schwarz:2002bc,He:2002zu} one finds, for example for $(I,J)=(i,j)$,
\begin{equation}
\begin{split}
\la n|H_3|\a_{0(r)}^{\dag\,k}\a_{0(s)}^{\dag\,l}|v\ra_{12}& \sim
\m\l'\frac{\sin^2 n\pi\b}{2\pi^2}\left(\d^{rs}+\frac{\sqrt{\a_r\a_s}}{\a_3}\right)S^{ijkl}\,,\\
\la n|H_3|\a_{k(r)}^{\dag\,k}\a_{-k(r)}^{\dag\,l}|v\ra_{12} & \sim
\m\l'\b(\b+1)\frac{\a_3}{\a_r}\frac{\sin^2 n\pi\b}{2\pi^2}S^{ijkl}\,,
\end{split}
\end{equation}
and the analogous expression for $(I,J)=(i',j')$ with an (inessential) overall minus sign.
Here
\begin{equation}
S^{ijkl}\equiv T^{ijkl}+\frac{1}{4}\d^{ij}T^{kl}\,,\quad
T^{ijkl}=\d^{ik}\d^{jl}+\d^{jk}\d^{il}-\frac{1}{2}\d^{ij}\d^{kl}\,,\quad
T^{kl} = -2\d^{kl}
\end{equation}
can be split into a symmetric-traceless and a trace part.
There is no contribution to the ${\bf 6}$ nor to $({\bf 4},{\bf 4})_{\pm}$. The sum over $k$ and the integral over $\b$ can
be done and the complete contribution of the impurity conserving channel with
bosonic excitations at one-loop is
\begin{equation}
\frac{\m g_2^2\l'}{4\pi^2}\frac{15}{64\pi^2n^2}\sum_{k,l}S^{ijkl}S^{ijkl}=
\frac{\m g_2^2\l'}{4\pi^2}\frac{15}{16\pi^2n^2}\Bigl[\frac{1}{2}\bigl(1+\frac{1}{2}\d^{ij}\bigr)+\frac{1}{4}\d^{ij}\Bigr]\,.
\end{equation}
The factors of $\frac{1}{2}(1+\frac{1}{2}\d^{ij})$ and $\frac{1}{4}\d^{ij}$
equal the normalization ${\mc N}$ of the string states.
Thus the contribution to the ${\bf 9}$ and ${\bf 1}$ is in both cases~\cite{Roiban:2002xr,Gomis:2003kj}
\begin{equation}\label{9-1}
\frac{\m g_2^2\l'}{4\pi^2}\frac{15}{16\pi^2n^2}\,.
\end{equation}
The second case with two fermionic oscillators in the double string state was not analyzed in~\cite{Roiban:2002xr,Gomis:2003kj}
and is quite tedious in the $SO(8)$ formalism. In the formulation used here the computation is comparable to the
one with bosonic oscillators.
This time the bosonic contribution including the normalization is for large $\m$
\begin{equation}
-\pi\a'\m^2\la n|K^M\K^N|E_a\ra
\sim\frac{\m\sqrt{\l'}}{2\pi}\b(\b+1)\sin^2n\pi\b\left(\d^{MI}\d^{NJ}+\d^{NI}\d^{MJ}\right)\,.
\end{equation}
This is automatically symmetric in $(I,J)$ and $(M,N)$. In particular,
only the {\em symmetric} part of $v_{mn}(Y)$ (and $v_{m'n'}(Y)$) can contribute.
In this case we have e.g.
\begin{align}
{}_{123}\la v|b_{-k(r)\,\b_1\b_2}b_{k(r)}^{\a_1\a_2}|E_b\ra & =
-\d^{\a_1}_{\b_1}\d^{\a_2}_{\b_2}e(k)Q^{rr}_{|k||k|}
\sim-\d^{\a_1}_{\b_1}\d^{\a_2}_{\b_2}\frac{k}{4\pi\bigl(\m\a_r\bigr)^2}\,,
\end{align}
and, therefore, there is no contribution to the representations ${\bf 9}$ and ${\bf 1}$ as the sum over $k$ is vanishing.
There is a contribution to
the representation $({\bf 4},{\bf 4})_+$:
the relevant matrix element with fermionic oscillators is ($k\ge0$)
\begin{equation}
-{}_{123}\la v|b_{k(r)\,\a_1\a_2}b_{-k(s)\,\da_1\da_2}Y^{\g_1\g_2}Z^{\dg_1\dg_2}|E_b\ra
=\d_{\a_1}^{\g_1}\d_{\a_2}^{\g_2}\d_{\da_1}^{\dg_1}\d_{\da_2}^{\dg_2}
\bar{G}_{k(r)}\bar{G}_{k(s)}\,.
\end{equation}
Taking the large $\m$ limit and summing all the contributions we find
\begin{equation}
\frac{\m g_2^2\l'}{4\pi^2}\frac{3}{8\pi^2 n^2}
-\frac{\m g_2^2\l'}{4n\pi^3}\int_{-1}^0 d\b\,\sin^4n\pi\b\cot n\pi\b(\b+1)
=\frac{\m g_2^2\l'}{4\pi^2}\frac{15}{16\pi^2 n^2}\frac{1}{2}\,,
\end{equation}
giving the same contribution as for ${\bf 9}$ and ${\bf 1}$.
\footnote{We have also checked this in the $SO(8)$ formalism.}

\subsection{Contribution of contact term}

To have a non-zero contribution from $Q_3^2$ the intermediate states need
to have an odd number of bosonic oscillators and an odd number of fermionic oscillators. Thus
the simplest contribution comes from the impurity conserving channel.
In this case the projector is
\begin{equation*}
\a_{0(1)}^{K\,\dag}b_{0(2)}^{a\,\dag}|v\ra\la v|b_{0(2)}^{a}\a_{0(1)}^{K}
+(1\leftrightarrow 2)
+\sum_{r,k\in\Zop}
\a_{k(r)}^{K\,\dag}b_{-k(r)}^{a\,\dag}|v\ra\la v|b_{-k(r)}^{a}\a_{k(r)}^{K}\,.
\end{equation*}
Including the normalization and taking into account that we already absorbed the string-length conserving delta-function $|Q_3\ra$ is
\begin{equation}
\eta|Q_{3\,\b_1\db_2}\ra \sim
-\sqrt{-\frac{\a'}{8\a_3^3}}\sqrt{-\b(\b+1)}\bigl(Y_{\b_1\g_2}\K^{\g_2}_{\db_1}-iZ_{\dg_1\b_2}\K^{\dg_1}_{\b_1}\bigr)|E_a\ra|E_b\ra\,,
\end{equation}
and the fermionic contribution is again rather trivial. The bosonic expectation value for large $\m$ has been
computed in~\cite{Roiban:2002xr} so we shall not repeat it here.
Taking into account all the contributions, doing the sum over $k$ and integral over $\b$ results in
\begin{equation}
\frac{1}{2}\frac{\m g_2^2\l'}{4\pi^2}\left(\frac{1}{12}+\frac{35}{32n^2\pi^2}\right)\,,
\end{equation}
for the antisymmetric ${\bf 6}$ and $({\bf 4},{\bf 4})_-$ and
\begin{equation}
\frac{\m g_2^2\l'}{4\pi^2}\left(\frac{1}{12}+\frac{5}{32n^2\pi^2}\right)
\Bigl[\frac{1}{2}\bigl(1+\frac{1}{2}\d^{ij}\bigr)+\frac{1}{4}\d^{ij}\Bigr]\,,
\end{equation}
for the ${\bf 1}$, ${\bf 9}$ and $({\bf 4},{\bf 4})_+$.
Summing the contributions of one-loop and contact diagrams
we see that all (bosonic) two-impurity irreducible representations of
$SO(4)\times SO(4)$ get the same contribution to the mass-shift from the
impurity-conserving channels
\begin{equation}
\d E_n^{(2)}=\frac{\m g_2^2\l'}{4\pi^2}\left(\frac{1}{12}+\frac{35}{32n^2\pi^2}\right)\,.
\end{equation}
This is in exact agreement with the gauge theory
result of~\cite{Beisert:2002bb,Constable:2002vq}.

\section{Conclusions}\label{section5}

In this paper we have presented a supersymmetric extension of the cubic interaction
vertex and dynamical supercharges with the manifest $SO(4)\times SO(4)$ symmetry
of the plane wave space-time. We have argued that this solution provides an
equivalent formulation to the previously known one in the $SO(8)$ formalism.
Our arguments were based on the discrete $\Zop_2$ symmetry of the plane wave, the equivalence of the effective
interaction vertices for states containing only bosonic creation oscillators in both formulations and, finally, the agreement
of the leading order mass-shift for the simplest string states.
It would be nice to find an exact proof of this conjectured equivalence and
understand the relevance of the solution recently given in~\cite{DiVecchia:2003yp}.

\vskip1cm
\section*{Acknowledgement}
\vskip0.2cm
I am grateful for discussions with G.~Arutyunov, N.~Beisert, J.~Plefka, M.~P\"{o}ssel and
S.~Theisen. In particular I would like to thank B.~Stefa\'nski
for useful comments on the manuscript.
This work was supported by GIF, the German-Israeli foundation
for Scientific Research.

\appendix

\section{Conventions and Notation}\label{appA}

The R-R flux in the plane wave geometry breaks the $SO(8)$ symmetry of the metric into
$SO(4)\times SO(4)\times\Zop_2$. Then
\begin{equation}
{\bf 8}_v \longrightarrow ({\bf 4},{\bf 1})\oplus ({\bf 1},{\bf 4})\,,\qquad
{\bf 8}_s \longrightarrow ({\bf 2},{\bf 2}) \oplus ({\bf 2'},{\bf 2'})\,,\qquad
{\bf 8}_c \longrightarrow ({\bf 2},{\bf 2'}) \oplus ({\bf 2'},{\bf 2})\,,
\end{equation}
where ${\bf 2}$ and ${\bf 2'}$ are the inequivalent Weyl representations of $SO(4)$.
We decompose $\g^I_{a\dot{a}}$ and $\g^I_{\dot{a}a}$ into $SO(4)\times SO(4)$ as follows
\begin{align}
\g^i_{a\dot{a}} & =
\begin{pmatrix}
0 & \s^i_{\a_1\db_1}\d_{\a_2}^{\b_2} \\ {\s^i}^{\da_1\b_1}\d^{\da_2}_{\db_2} & 0
\end{pmatrix}\,,\qquad
\g^i_{\dot{a}a} =
\begin{pmatrix}
0 & \s^i_{\a_1\db_1}\d^{\da_2}_{\db_2} \\ {\s^i}^{\da_1\b_1}\d_{\a_2}^{\b_2} & 0
\end{pmatrix}\,,\\
\g^{i'}_{a\dot{a}} & =
\begin{pmatrix}
-\d_{\a_1}^{\b_1}\s^{i'}_{\a_2\db_2} & 0 \\ 0 & \d^{\da_1}_{\db_1}{\s^{i'}}^{\da_2\b_2}
\end{pmatrix}\,,\qquad
\g^{i'}_{\dot{a}a} =
\begin{pmatrix}
-\d_{\a_1}^{\b_1}{\s^{i'}}^{\da_2\b_2} & 0 \\ 0 & \d^{\da_1}_{\db_1}\s^{i'}_{\a_2\db_2}
\end{pmatrix}\,.
\end{align}
Here the $\s$-matrices consist of the usual Pauli-matrices together with the 2d unit matrix
\begin{equation}
\s^i_{\a\da}=\bigl(i\t^1,i\t^2,i\t^3,-1\bigr)_{\a\da}
\end{equation}
and we raise and lower spinor indices with the two-dimensional Levi-Civita symbols, e.g.
\begin{equation}
\s^i_{\a\da} = \e_{\a\b}\e_{\da\db}{\s^i}^{\db\b}
\equiv \e_{\a\b}{\s^i}^{\b}_{\da} \equiv \e_{\da\db}{\s^i}^{\db}_{\a}\,.
\end{equation}
The $\s$-matrices obey the relations
\begin{equation}
\s^i_{\a\da}{\s^j}^{\da\b}+\s^j_{\a\da}{\s^i}^{\da\b}
=2\d^{ij}\d_{\a}^{\b}\,,\qquad
{\s^i}^{\da\a}\s^j_{\a\db}+{\s^j}^{\da\a}\s^i_{\a\db}=
2\d^{ij}\d^{\da}_{\db}\,.
\end{equation}
In particular, in this basis
\begin{equation}
\Pi_{ab} =
\begin{pmatrix} \bigl(\s^1\s^2\s^3\s^4\bigr)_{\a_1}^{\b_1}\d_{\a_2}^{\b_2} & 0 \\
0 & \bigl(\s^1\s^2\s^3\s^4\bigr)^{\da_1}_{\db_1}\d^{\da_2}_{\db_2}
\end{pmatrix} =
\begin{pmatrix}
\d_{\a_1}^{\b_1}\d_{\a_2}^{\b_2} & 0 \\ 0 & -\d^{\da_1}_{\db_1}\d^{\da_2}_{\db_2}
\end{pmatrix}\,,
\end{equation}
that is $(1\pm\Pi)/2$ project onto $({\bf 2},{\bf 2})$ and
$({\bf 2'},{\bf 2'})$, respectively.
A number of  useful relations we have to use are
\begin{align}
\label{rel1}
\e_{\a\b}\e^{\g\d} & = \d_{\a}^{\d}\d_{\b}^{\g}-\d_{\a}^{\g}\d_{\b}^{\d}\,,\\
\s^i_{\a\db}{\s^j}^{\db}_{\b} & = -\d^{ij}\e_{\a\b}+\s^{ij}_{\a\b}\,,\qquad
(\s^{ij}_{\a\b}\equiv \s^{[i}_{\a\da}{\s^{j]}}^{\da}_{\b}=\s^{ij}_{\b\a})\\
\s^i_{\a\da}{\s^j}^{\a}_{\db} & = -\d^{ij}\e_{\da\db}+\s^{ij}_{\da\db}\,,
\qquad (\s^{ij}_{\da\db}\equiv \s^{[i}_{\a\da}{\s^{j]}}^{\a}_{\db}=\s^{ij}_{\db\da})\\
\s^k_{\a\da}\s^{k}_{\b\db} & = 2\e_{\a\b}\e_{\da\db}\,,\\
\s^{kl}_{\a\b}\s^{kl}_{\g\d} & = 4(\e_{\a\g}\e_{\b\d}+\e_{\a\d}\e_{\b\g})\,,\\
\s^{kl}_{\a\b}\s^{kl}_{\dg\dd} & = 0\,,\\
\label{rel7}
2\s^i_{\a\da}\s^{j}_{\b\db} & = \d^{ij}\e_{\a\b}\e_{\da\db}
+\s^{k(i}_{\a_1\b_1}\s^{j)k}_{\da_1\db_1}
-\e_{\a\b}\s^{ij}_{\da\db}-\s^{ij}_{\a\b}\e_{\da\db}\,.
\end{align}

\section{Useful identities and (anti)commutators}\label{appB}

The multi-linears in $Y$, defined in section~\ref{section3},~\eqref{y2}--\eqref{y4}
satisfy
\begin{align}
\label{id1}
Y_{\a_1\a_2}Y_{\b_1\b_2} & = -\frac{1}{2}
\bigl(\e_{\a_1\b_1}Y^2_{\a_2\b_2}+\e_{\a_2\b_2}Y^2_{\a_1\b_1}\bigr)\,,\\
\label{id2}
Y_{\a_1\a_2}Y^2_{\b_2\g_2} & = -\frac{1}{3}\bigl(
\e_{\a_2\g_2}Y^3_{\a_1\b_2}+\e_{\a_2\b_2}Y^3_{\a_1\g_2}\bigr)\,,\\
\label{id3}
Y_{\a_1\a_2}Y^2_{\b_1\g_1} & = \frac{1}{3}\bigl(
\e_{\a_1\b_1}Y^3_{\g_1\a_2}+\e_{\a_1\g_1}Y^3_{\a_1\a_2}\bigr)\,,\\
\label{id4}
Y^3_{\b_1\g_2}Y_{\a_1\d_2} & = \frac{1}{4}\e_{\b_1\a_1}\e_{\g_2\d_2}Y^4\,,
\end{align}
and analogously for $Z$. Then the most general ansatz of the form
in~\eqref{ansatz1} is
\begin{align}\label{ansatz2}
f^i_{\b_1\db_2}\K^i & =
\Bigl(Z\bigl[(a_{0,1}+a_{4,1}Y^4)\e+a_{2,1}Y^2\bigr]
+Z^3\bigl[(a_{0,3}+a_{4,3}Y^4)\e+a_{2,3}Y^2\bigr]
\Bigr)_{\dg_1\db_2;\b_1\g_1}\K^{\dg_1\g_1}\,,\\
g^{i'}_{\b_1\db_2}\K^{i'} & =
\Bigl(Y\bigl[(a_{1,0}+a_{1,4}Z^4)\e+a_{1,2}Z^2\bigr]
+Y^3\bigl[(a_{3,0}+a_{3,4}Z^4)\e+a_{3,2}Z^2\bigr]
\Bigr)_{\b_1\g_2;\db_2\dg_2}\K^{\dg_2\g_2}\,.
\end{align}
The (anti)commutators of the free supercharges with the prefactor
constituents in the $SO(4)\times SO(4)$ basis are (for simplicity we suppress factors of $\eta\sqrt{\frac{-2\a}{\a'}}$).
\begin{align}
\label{ac1Y}
\sum_r\{Q_{(r)\,\a_1\da_2},Y_{\b_1\b_2}\} & = -i\e_{\b_1\a_1}K_{\b_2\da_2},
&\sum_r\{Q_{(r)\,\da_1\a_2},Y_{\b_1\b_2}\} & = iK_{\b_1\da_1}\e_{\b_2\a_2},\\
\label{ac1Z}
\sum_r\{Q_{(r)\,\a_1\da_2},Z_{\db_1\db_2}\} & = K_{\a_1\db_1}\e_{\da_2\db_2},
&\sum_r\{Q_{(r)\,\da_1\a_2},Z_{\db_1\db_2}\} & = \e_{\da_1\db_1}K_{\a_2\db_2}.
\end{align}
\begin{align}
\label{c1K}
\sum_r[Q_{(r)\,\a_1\da_2},\K_{\b_1\db_1}] & = -\frac{4\m\a}{\a'}i\e_{\b_1\a_1}Z_{\db_1\da_2},
&\sum_r[Q_{(r)\,\da_1\a_2},\K_{\b_1\db_1}] & = \frac{4\m\a}{\a'}\e_{\db_1\da_1}Y_{\b_1\a_2},\\
\label{c2K}
\sum_r[Q_{(r)\,\a_1\da_2},\K_{\b_2\db_2}] & = \frac{4\m\a}{\a'}\e_{\da_2\db_2}Y_{\a_1\b_2},
&\sum_r[Q_{(r)\,\da_1\a_2},\K_{\b_2\db_2}] & = \frac{4\m\a}{\a'}i\e_{\a_2\b_2}Z_{\da_1\db_2}.
\end{align}
\begin{align}
\label{qv1}
\sum_rQ_{(r)\,\a_1\da_2}|V\ra & =
-\left(K_{\da_2}^{\a_2}Y_{\a_1\a_2}+iK_{\a_1}^{\da_1}Z_{\da_1\da_2}\right)|V\ra\,,\\
\label{qv2}
\sum_rQ_{(r)\,\da_1\a_2}|V\ra & =
\left(K_{\da_1}^{\a_1}Y_{\a_1\a_2}-iK_{\a_2}^{\da_2}Z_{\da_1\da_2}\right)|V\ra\,.
\end{align}
Further useful relations are
\begin{equation}\label{1Y234}
\begin{split}
\sum_r[Q_{(r)\,\a_1\da_2},Y^2_{\b_1\g_1}] & =
-i\bigl(\e_{\a_1\b_1}Y_{\g_1\g_2}+\e_{\a_1\g_1}Y_{\b_1\g_2}\bigr)K^{\g_2}_{\da_2}\,,\\
\sum_r[Q_{(r)\,\a_1\da_2},Y^2_{\b_2\g_2}] & =
i\bigl(Y_{\a_1\b_2}K_{\g_2\da_2}+Y_{\a_1\g_2}K_{\b_2\da_2}\bigr)\,,\\
\sum_r\{Q_{(r)\,\a_1\da_2},Y^3_{\b_1\g_2}\} & =
3iY_{\a_1\g_2}Y_{\b_1\d_2}K^{\d_2}_{\da_2}\,,\\
\sum_r[Q_{(r)\,\a_1\da_2},Y^4] & =
4iY^3_{\a_1\a_2}K^{\a_2}_{\da_2}\,,
\end{split}
\end{equation}
\begin{equation}\label{2Y234}
\begin{split}
\sum_r[Q_{(r)\,\db_1\b_2},Y^2_{\a_1\g_1}] & =
-i\bigl(Y_{\g_1\b_2}K_{\a_1\db_1}+Y_{\a_1\b_2}K_{\g_1\db_1}\bigr)\,,\\
\sum_r[Q_{(r)\,\db_1\b_2},Y^2_{\a_2\g_2}] & =
-i\bigl(\e_{\a_2\b_2}Y_{\g_1\g_2}+\e_{\g_2\b_2}Y_{\g_1\a_2}\bigr)K^{\g_1}_{\db_1}\,,\\
\sum_r\{Q_{(r)\,\db_1\b_2},Y^3_{\a_1\g_2}\} & =
3iY_{\a_1\b_2}Y_{\g_1\g_2}K^{\g_1}_{\db_1}\,,\\
\sum_r[Q_{(r)\,\db_1\b_2},Y^4] & =
-4iY^3_{\a_1\b_2}K^{\a_1}_{\db_1}\,,
\end{split}
\end{equation}
\begin{equation}\label{1Z234}
\begin{split}
\sum_r[Q_{(r)\,\a_1\da_2},Z^2_{\db_1\dg_1}] & =
K_{\a_1\db_1}Z_{\dg_1\da_2}+K_{\a_1\dg_1}Z_{\db_1\da_2}\,,\\
\sum_r[Q_{(r)\,\a_1\da_2},Z^2_{\db_2\dg_2}] & =
-\bigl(\e_{\da_2\db_2}Z_{\dg_1\dg_2}+\e_{\da_2\dg_2}Z_{\dg_1\db_2}\bigr)K_{\a_1}^{\dg_1}\,,\\
\sum_r\{Q_{(r)\,\a_1\da_2},Z^3_{\db_1\dg_2}\} & =
-3Z_{\db_1\da_2}Z_{\dg_1\dg_2}K^{\dg_1}_{\a_1}\,,\\
\sum_r[Q_{(r)\,\a_1\da_2},Z^4] & =
4Z^3_{\da_1\da_2}K^{\da_1}_{\a_1}\,,
\end{split}
\end{equation}
\begin{equation}\label{2Z234}
\begin{split}
\sum_r[Q_{(r)\,\db_1\b_2},Z^2_{\da_1\dg_1}] & =
-\bigl(\e_{\db_1\da_1}Z_{\dg_1\dg_2}+\e_{\db_1\dg_1}Z_{\da_1\dg_2}\bigr)K_{\b_2}^{\dg_2}\,,\\
\sum_r[Q_{(r)\,\db_1\b_2},Z^2_{\da_2\dg_2}] & =
K_{\b_2\da_2}Z_{\db_1\dg_2}+K_{\b_2\dg_2}Z_{\db_1\da_2}\,,\\
\sum_r\{Q_{(r)\,\db_1\b_2},Z^3_{\da_1\dg_2}\} & =
3Z_{\db_1\dg_2}Z_{\da_1\dd_2}K^{\dd_2}_{\b_2}\,,\\
\sum_r[Q_{(r)\,\db_1\b_2},Z^4] & =
4Z^3_{\db_1\dg_2}K^{\dg_2}_{\b_2}\,.
\end{split}
\end{equation}
All the (anti)commutators involving $\wt{Q}_{(r)}$ can be obtained with the help of
the reality properties.

\section{More detailed computations}\label{appC}

In this section we provide some details for the computation determining
$|Q_{3\,\b_1\db_1}\ra$ and $|H_3\ra$, as well as for the proof of~\eqref{dyn3}.
We substitute the general ansatz given in~\eqref{ansatz1} and~\eqref{ansatz2}
in~\eqref{1}, use the various identities~\eqref{rel1}--\eqref{rel7},~\eqref{id1}--\eqref{id4} and
(anti)commutators~\eqref{ac1Y}--\eqref{2Z234} given in the previous
sections and analyze the resulting contributions order
by order in $Y$ and $Z$.

Consider first the terms independent of $K$, $\K$. We get (here and below we
suppress an overall factor of $\bar{\eta}\sqrt{\frac{\a'}{-2\a}}$)
\begin{equation}
-4\frac{\m\a}{\a'}\e_{\a_1\b_1}\e_{\da_2\db_2}
\Bigl(a_{3,0}Y^4+ia_{0,3}Z^4+(a_{3,4}+ia_{4,3})Y^4Z^4\Bigr)\,,\quad
\text{and}\quad a_{2,1}=ia_{1,2}\,.
\end{equation}
The latter condition arises to cancel a contribution proportional to the tensor $Y^2_{\a_1\b_1}Z^2_{\da_2\db_2}$.
Next we take into account the terms proportional to $K\K$ and summarize the results order by
order in fermions in the table below: 
\begin{center}
\begin{tabular}{|c|c|c|} \hline
${\mc O}(Y^mZ^n)$ & contribution & condition \\ \hline
(0,0) &
$-2\e_{\a_1\b_1}\e_{\da_2\db_2}\bigl(a_{0,1}K^i\K^j\d^{ij}+ia_{1,0}K^{i'}\K^{j'}\d^{i'j'}\bigr)$
& --- \\ \hline
(2,0) &
$\e_{\a_1\b_1}\e_{\da_2\db_2}\bigl(a_{2,1}K^i\K^j{Y^2}^{ij}-a_{1,0}K^{i'}\K^{j'}{Y^2}^{i'j'}\bigr)$ &
$a_{3,0}=\frac{i}{3}a_{1,0}$ \\ \hline
(0,2) &
$-i\e_{\a_1\b_1}\e_{\da_2\db_2}\bigl(a_{0,1}K^i\K^j{Z^2}^{ij}-a_{1,2}K^{i'}\K^{j'}{Z^2}^{i'j'}\bigr)$ &
$a_{0,3}=\frac{i}{3}a_{0,1}$ \\ \hline
(1,1) &
$2a_{0,1}\e_{\a_1\b_1}\e_{\da_2\db_2}
\bigl(K_{\g_1\dg_1}\K_{\g_2\dg_2}+\K_{\g_1\dg_1}K_{\g_2\dg_2}\bigr)Y^{\g_1\g_2}Z^{\dg_1\dg_2}$ &
$a_{1,2}=a_{0,1}=-ia_{1,0}$ \\ \hline
(4,0) &
$-2\e_{\a_1\b_1}\e_{\da_2\db_2}Y^4
\bigl(a_{4,1}K^i\K^j\d^{ij}+\frac{1}{12}a_{0,1}K^{i'}\K^{j'}\d^{i'j'}\bigr)$ &
--- \\ \hline
(0,4) &
$-2\e_{\a_1\b_1}\e_{\da_2\db_2}Z^4
\bigl(\frac{1}{12}a_{0,1}K^i\K^j\d^{ij}+ia_{1,4}K^{i'}\K^{j'}\d^{i'j'}\bigr)$ &
--- \\ \hline
(2,2) &
$-\frac{a_{0,1}}{2}\e_{\a_1\b_1}\e_{\da_2\db_2}
\bigl(K^i\K^j\bigl(Y^2Z^2\bigr)^{ij}-K^{i'}\K^{j'}\bigl(Y^2Z^2\bigr)^{i'j'}\bigr)$ &
$a_{3,2}=-ia_{2,3}=\frac{i}{3}a_{0,1}$ \\ \hline
(1,3) &
$-\frac{2}{3}a_{1,0}\e_{\a_1\b_1}\e_{\da_2\db_2}
\bigl(K_{\g_1\dg_1}\K_{\g_2\dg_2}-\K_{\g_1\dg_1}K_{\g_2\dg_2}\bigr)
Y^{\g_1\g_2}{Z^3}^{\dg_1\dg_2}$ &
$a_{1,4}=-\frac{1}{12}a_{1,0}$ \\ \hline
(3,1) &
$\frac{2}{3}a_{1,0}\e_{\a_1\b_1}\e_{\da_2\db_2}
\bigl(K_{\g_1\dg_1}\K_{\g_2\dg_2}-\K_{\g_1\dg_1}K_{\g_2\dg_2}\bigr)
{Y^3}^{\g_1\g_2}Z^{\dg_1\dg_2}$ &
$a_{4,1}=-\frac{i}{12}a_{1,0}$ \\ \hline
(4,2) &
$-\frac{a_{1,0}}{12}\e_{\a_1\b_1}\e_{\da_2\db_2}Y^4
\bigl(K^i\K^j\bigl(Z^2\bigr)^{ij}+K^{i'}\K^{j'}\bigl(Z^2\bigr)^{i'j'}\bigr)$ &
$a_{4,3}=\frac{i}{36}a_{0,1}$ \\ \hline
(2,4) &
$\frac{a_{1,0}}{12}\e_{\a_1\b_1}\e_{\da_2\db_2}Z^4
\bigl(K^i\K^j\bigl(Y^2\bigr)^{ij}+K^{i'}\K^{j'}\bigl(Y^2\bigr)^{i'j'}\bigr)$ &
$a_{3,4}=-\frac{i}{36}a_{1,0}$ \\ \hline
(3,3) &
$\frac{2}{9}a_{0,1}\e_{\a_1\b_1}\e_{\da_2\db_2}
\bigl(K_{\g_1\dg_1}\K_{\g_2\dg_2}+\K_{\g_1\dg_1}K_{\g_2\dg_2}\bigr)
{Y^3}^{\g_1\g_2}{Z^3}^{\dg_1\dg_2}$ &
--- \\ \hline
(4,4) &
$-\frac{1}{72}a_{0,1}\e_{\a_1\b_1}\e_{\da_2\db_2}Y^4Z^4
\bigl(K^i\K^j\d^{ij}-K^{i'}\K^{j'}\d^{i'j'}\bigr)$ &
--- \\ \hline
\end{tabular}
\end{center}
Two further relations we had to use in the above are e.g.
\begin{equation}
2Y_{[\a_1\g_2}K^{\dg_1}_{\b_1]} = \e_{\a_1\b_1}Y^{\g_1}_{\g_2}K^{\dg_1}_{\g_1}\,,
\end{equation}
and
\begin{align}
&Y_{\a_1\g_2}Z_{\dg_1\da_2}K^{\dg_1}_{\b_1}\K^{\g_2}_{\db_2}
+Y_{\b_1\g_2}Z_{\dg_1\db_2}K^{\dg_1}_{\a_1}\K^{\g_2}_{\da_2}
-Y_{\a_1\g_2}Z_{\dg_1\db_2}K^{\dg_1}_{\b_1}\K^{\g_2}_{\da_2}
-Y_{\b_1\g_2}Z_{\dg_1\da_2}K^{\dg_1}_{\a_1}\K^{\g_2}_{\db_2}\nonumber\\
&=\e_{\a_1\b_1}\e_{\da_2\db_2}Y^{\g_1\g_2}Z^{\dg_1\dg_2}K_{\g_1\dg_1}\K_{\g_2\dg_2}\,.
\end{align}
The computation determining $|Q_{3\,\db_1\b_2}\ra$ proceeds in an analogous fashion.
One can check that then~\eqref{3} is identically satisfied.

The equations arising from the constraint~\eqref{dyn3} can again be proved by
associating phase factors to the oscillators as in~\cite{Green:1983hw,Pankiewicz:2002tg}
and integrating by parts. Recall that we
associate the world-sheet coordinate dependence with the oscillators as
\begin{equation}\label{trick}
\begin{pmatrix} b_{n(r)} \\ b_{-n(r)} \end{pmatrix}
\longrightarrow e^{-i\o_{n(r)}\t/\a_r}
\begin{pmatrix}
\cos\frac{n\s_r}{\a_r} & -\sin\frac{n\s_r}{\a_r} \\
\sin\frac{n\s_r}{\a_r} & \cos\frac{n\s_r}{\a_r}
\end{pmatrix}
\begin{pmatrix} b_{n(r)} \\ b_{-n(r)} \end{pmatrix}\,,
\end{equation}
and analogously for the bosonic oscillators. We also define $\p_{\s}\equiv\sum_r\p_{\s_r}$.
One needs
\begin{equation}
\begin{split}
\sum_r[Q_{(r)\,\a_1\da_2},K_{\b_1\db_1}] & =
-4\frac{\a}{\a'}\e_{\b_1\a_1}\bigl(\p_{\t}+\p_{\s}\bigr)Z_{\db_1\da_2}\,,\\
\sum_r[Q_{(r)\,\da_1\a_2},K_{\b_1\db_1}] & =
-4i\frac{\a}{\a'}\e_{\db_1\da_1}\bigl(\p_{\t}+\p_{\s}\bigr)Y_{\b_1\a_2}\,,
\end{split}
\end{equation}
\begin{equation}
\begin{split}
\sum_r[Q_{(r)\,\a_1\da_2},K_{\b_2\db_2}] & =
-4i\frac{\a}{\a'}\e_{\da_2\db_2}\bigl(\p_{\t}+\p_{\s}\bigr)Y_{\a_1\b_2}\,,\\
\sum_r[Q_{(r)\,\da_1\a_2},K_{\b_2\db_2}] & =
4\frac{\a}{\a'}\e_{\a_2\b_2}\bigl(\p_{\t}+\p_{\s}\bigr)Z_{\da_1\db_2}\,,
\end{split}
\end{equation}
and
\begin{equation}\label{dV}
\p_{\s}|V\ra = -\frac{i}{4}
\left[\frac{\a'}{\a}\bigl(K^IK^I-\K^I\K^I\bigr)-
4\bigl(Y_{\a_1\a_2}\p_{\t}Y^{\a_1\a_2}-Z_{\da_1\da_2}\p_{\t}Z^{\da_1\da_2}\bigr)\right]|V\ra\,.
\end{equation}
Notice that $\p_{\t}\to-\p_{\t}$ under complex conjugation. Further identities include e.g.
\begin{align}
\p_{\s}Y^3_{\a_1\b_2} & = \frac{3}{2}
\bigl(Y^2_{\a_1\g_1}\p_{\s}Y^{\g_1}_{\b_2}-Y^2_{\b_2\g_2}\p_{\s}Y^{\g_2}_{\a_1}\bigr)\,,\\
\p_{\s}Y^4 & = 4Y^3_{\a_1\a_2}\p_{\s}Y^{\a_1\a_2}\,,\\
Y^2_{\a_1\b_1}Y_{\g_1\g_2}\p_{\s}Y^{\g_1\g_2} & = -\frac{1}{3}
\bigl(Y^3_{\a_1\g_2}\p_{\s}Y^{\g_2}_{\b_1}+Y^3_{\b_1\g_2}\p_{\s}Y^{\g_2}_{\a_1}\bigr)\,,\\
Y^2_{\a_2\b_2}Y_{\g_1\g_2}\p_{\s}Y^{\g_1\g_2} & = \frac{1}{3}
\bigl(Y^3_{\g_1\a_2}\p_{\s}Y^{\g_1}_{\b_2}+Y^3_{\g_1\b_2}\p_{\s}Y^{\g_1}_{\a_2}\bigr)\,,\\
Y ^4\p_{\s}Y_{\a_1\a_2} & = -\p_{\s}\bigl(Y^4Y_{\a_1\a_2}\bigr)=0\,.
\end{align}

\section{Functional expressions for the prefactor}\label{appD}

The functional expressions for the fermionic constituents of the prefactor are
\begin{align}
Y(\s) & \equiv -2\sqrt{2}\pi
(\pi\a_1-\s)^{1/2}\bigl(\l_1(\s)+\l_1(-\s)\bigr)\,,\\
Z(\s) & \equiv -\frac{1}{\sqrt{2}}(\pi\a_1-\s)^{1/2}\bigl(\vt_1(\s)+\vt_1(-\s)\bigr)\,.
\end{align}
These satisfy
\begin{equation}
\lim_{\s\to\pi\a_1}Y(\s)|V\ra = Y|V\ra\,,\qquad \lim_{\s\to\pi\a_1}Z(\s)|V\ra = Z|V\ra\,.
\end{equation}
We show this for $Y(\s)$, the analysis involving $Z(\s)$ is similar. We have
\begin{align}
\lim_{\s\to\pi\a_1}Y(\s)|V\ra & =-\frac{2}{\sqrt{\a_1}}
\lim_{\e\to0}\e^{1/2}\sum_{n=1}^{\infty}(-1)^n\cos(n\e/\a_1)\times \nonumber\\
&\times\left[\sqrt{2}\frac{1+\Pi}{2}\L \bar{Q}^1_n+
\sum_{r=1}^3\sum_{m=1}^{\infty}Q^{1r}_{nm}b^{\dag}_{m(r)}\right]|V\ra\,.
\end{align}
For large $n$
\begin{align}
\bar{Q}_n^1 & \sim \frac{1}{\sqrt{\a_1}}\bigl(C\bar{N}^1\bigr)_n\,,\\
Q^{1r}_{nm} & \sim \sqrt{\frac{\a'}{\a_1}}\left(
\frac{1+\Pi}{2}\bigl(C\bar{N}^1\bigr)_n\bar{G}_{m(r)}
-\frac{1-\Pi}{2}\frac{\a_1}{\a_r}\bar{N}^1_n\bigl(C\bar{G}_{(r)}\bigr)_m\right)\,.
\end{align}
In fact, only the terms proportional to $(1+\Pi)$ survive the limit $\e\to 0$. Then we get
\begin{equation}
\lim_{\s\to\pi\a_1}Y(\s)|V\ra = f(\m)(1-4\m\a K)^{-1/2}Y|V\ra\,,
\end{equation}
where
\begin{equation}
f(\m)\equiv-2\frac{\sqrt{-\a}}{\a_1}\lim_{e\to0}\e^{1/2}
\sum_{n=1}^{\infty}(-1)^nn\cos(n\e/\a_1)\bar{N}^1_n=(1-4\m\a K)^{1/2}.
\end{equation}
The last identity was conjectured in~\cite{Pankiewicz:2002tg} and recently proved in~\cite{He:2002zu}.

\providecommand{\href}[2]{#2}\begingroup\raggedright\endgroup


\end{document}